# Silicon Atomic Quantum Dots Enable Beyond-CMOS Electronics

Robert A. Wolkow[1,2,3], Lucian Livadaru[3] Jason Pitters[2], Marco Taucer[3], Paul Piva[3], Mark Salomons[2], Martin Cloutier[2], Bruno V.C. Martins[1]

[1] Department of Physics, University of Alberta,
11322-89 Avenue, Edmonton, Alberta, T6G 2G7,
rwolkow@ualberta.ca,
[2] National Institute for Nanotechnology, National Research Council of Canada,
11421 Saskatchewan Drive, Edmonton, Alberta, T6G 2M9, Canada
[3] Quantum Silicon Inc., 11421 Saskatchewan Drive, AB, T6G 2M9 Edmonton AB, Canada

**Abstract.** We review our recent efforts in building atom-scale quantum-dot cellular automata circuits on a silicon surface. Our building block consists of silicon dangling bond on a H-Si(001) surface, which has been shown to act as a quantum dot. First the fabrication, experimental imaging, and charging character of the dangling bond are discussed. We then show how precise assemblies of such dots can be created to form artificial molecules. Such complex structures can be used as systems with custom optical properties, circuit elements for quantum-dot cellular automata, and quantum computing. Considerations on macro-to-atom connections are discussed.

## 1 Preliminaries

There are two broad problems facing any prospective nano-scale electronic device building block. It must have an attractive property such as to switch, store or conduct information, but also, there must be an established architecture in which the new entity can be deployed and wherein it will function in concert with other elements. Nanoscale electronic device research has in few instances so far led to functional blocks that are ready for insertion into existing device designs. In this work we discuss a range of atom-based device concepts which, while requiring further development before commercial products can emerge, have the great advantage that an overall architecture is well established that calls for exactly the type of building block we have developed.

The atomic silicon quantum dot (ASiQD) described here fits within ultra low power schemes for beyond CMOS electronics based upon quantum dots that have been refined over the past 2 decades. The well known quantum dot cellular automata (QCA) scheme due to Lent and co-workers [1, 2] achieves classical binary logic functions without the use of conventional current-based technology.

Within this scheme, the binary states "1" and "0" are encoded in the position of electric charge. Variants exist but most commonly the basic cell consist of a square (or rectangular) quantum dot ensemble occupied by 2 electrons. Electrons freely tunnel among the quantum dots in a cell, while electron tunneling between cells does not occur. Within a cell, two classically equivalent states exist, each with electrons placed on the diagonal of the cell. Multiple cells couple and naturally mimic the electron configuration of nearest-neighbour cells. In general, cell-cell interactions must be described quantum mechanically but to a good approximation they are described simply by electrostatic interactions. A line of coupled cells serves as a binary wire. When a terminal cell is forced by a nearby electrode to be in one of its two polarized states, adjacent cells copy that configuration to transfer that input state to the other terminus. This transfer can happen spontaneously or can be zonally regimented by a clock signal that controls inter-dot barriers, or some other parameter. The last key feature of QCA is that three binary lines acting as computation inputs and one line acting as output can converge on a node cell to create a majority gate. If two of the three input lines are of one binary state, the fourth side of the node cell will output the majority state. Variants of such an arrangement allow for the realization of a full logical basis. To date, all manner of digital circuits have been designed, from memories to multipliers to even a microprocessor.

While complex working circuits have not yet been realized, all the rudimentary circuit elements have been already experimentally demonstrated [3, 4]. Furthermore, the input state of a QCA circuit has been externally controlled and the output has been successfully read-out by a coupled single electron transistor [5, 6]. Until the present work, all available quantum dots, typically consisting of thousands of atoms, had narrowly spaced energy levels requiring ultra-low temperature to exhibit desired electronic properties. Moreover, approximately as many wires as quantum dots were required to adjust electron filling, a scenario that would greatly limit the complexity of circuitry that could be explored.

A prospect for highly complex and room temperature operational QCA circuitry suddenly emerged with the discovery of atomic silicon quantum dots. Figure 1 shows a schematic 4-dot QCA cell on the left occupied with two electrons (indicated by blackened circles). On the right is an STM image of a real atom-scale cell made of 4 ASiQDs, the cell being less than 2 nm on a side. The darker of the two dots are predominantly electron occupied.

The ultimate small size of the ASiQDs leads to ultimate wide spacing of energy levels indeed sufficiently widely spaced to allow room temperature device operation. The ASiQDs can be prepared in a native 1- charge state (charge is expressed in elementary charge units henceforth). Close placement of dots causes Coulombic repulsion and even removal of an electron to the silicon substrate conduction band. By fabricating dots at an appropriate spacing, a desired level of electron occupation can be predetermined, eliminating the need for many wires. As all atomic dots are identical, and their placement occurs in exact registry with the regular atomic structure of the underlying crystalline lattice, structures with uniquely homogeneous and reproducible characteristics can be

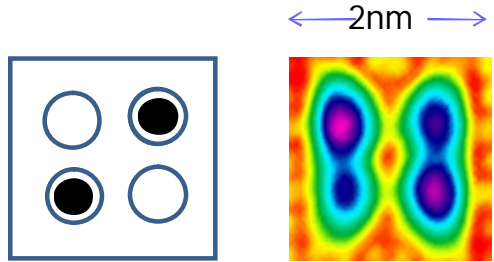
**Fig. 1.** Left: schematic representation of a square QCA cell with 2 electrons (blackened circles) positioned on the diagonal configuration. Right: STM map of an actual ASiQD structure with 4 dots in a square pattern as an embodiment of the QCA cell on the left. Electron population is predominant on the same diagonal as indicated on the schematic on the left.

in principle fabricated. A further advantage lies in the fact that these dots are entirely made of and upon silicon, enabling compatibility with silicon CMOS circuitry. This allows the merging of established and new technologies, greatly easing the path to deployment.

Challenges in the precise positioning of single silicon atom dots previously limited creation of more than a 4-atom ensemble. New developments have enabled patterns with hundreds of atoms to be fabricated with error rates close to those required for functioning computation circuit elements. A path to further improvements appears to be in hand.

Information storage, transfer and computation without use of conventional electrical current, with several orders lower power consumption than CMOS appear within reach. Prospects for extremely small size and weight appear good, too, as are those for extreme speed. Existing true 2-dimensional circuit layouts indicate a great reduction in the need for multilayer interconnects. The all-silicon aspect of this approach leads to a natural CMOS compatibility and therefore an early entry point via a hybrid CMOS-ASiQD technology. Room temperature as opposed to cryo operation is very attractive. The materials stability of the system up to $> 200^oC$ is comparable to conventional electronics. Furthermore, the possibility of deployment in an analog mode broadens the appeal and power of the approach. A natural ability to merge with Si-based sensor circuitry is desirable too. As discussed below, potential applications in quantum information are also very appealing.

## 2 Preparing and visualizing silicon surface dangling bonds

A silicon dangling bond, DB, exists at a silicon atom that is under-coordinated, that is where a silicon atom has only 3, rather the regular 4 bonding partners. In this discussion we will focus on the DB on the hydrogen terminated (100) face of a silicon crystal, abbreviated H-Si(100). Atomically flat, ordered H-termination is ordinarily achieved by cracking Hydrogen gas, $H_2$, into H atoms by collision with a hot tungsten filament and allowing those H atoms to react with a clean silicon surface in a vacuum chamber. If the H-termination process is incomplete, or if an H atom is removed by some chemical or physical means, a DB is created. H atoms removed by the local action of a scanning tunneling microscopy (STM) are the focus here. Broadly speaking, the scanning motion of the tip can be halted to direct an intense electrical current in the vicinity of a single Si-H surface bond [7–10]. At approximately a 2 Volts bias between tip and sample it is understood that multiple vibrational excitations lead to dissociation of the Si-H bond. At near 5 Volts bias it is thought the Si-H bond can be excited to a dissociative state, as in a photochemical bond breaking event. Other not well understood factors are at play, such as a catalytic effect, intimately depending on particular tip apex structure and composition that might ease the Si-H bond apart as a substantial H atom-tip bond forms while the Si-H bond lengthens and weakens. The fate of removed H atoms is unclear though there is substantial evidence, in the form of H atom donation to the surface that some atoms reside on the STM tip [11, 12]. Many details related to exact position of the tip and precise metering of the energetic bond breaking process so as to create just the change desired and not other surface alterations will be touched upon in the section on Quantum Silicon Incorporated and the commercial drive to fabricate atom scale silicon devices.

Figure 2a shows a model of a H-Si(100) surface. Silicon atoms are yellow. Hydrogen atoms are white. Note the surface silicon atoms are combined with H atoms in a 1 to 1 ratio. Note also that the surface silicon atoms deviate from the bulk structure not only in that they have H atom partners, but also in that each surface silicon atom is paired-up into a dimer unit. The dimers exist in rows.

Figure 2b shows a constant current STM image of a H-Si(100) surface. The dimer units are 3.84 Angstroms separated along a dimer row. The rows are separated by twice that distance, 7.68 Angstroms. The overlaid grid of black bars marks the position of the silicon surface dimer bonds. To reiterate, there is an H atom positioned at both ends of each dimer unit.

Figure 3 indicates the localized creation of a dangling bond upon action directed by a scanned probe tip. Figure 3b shows an STM image of several DBs so created [13].

## 3 The Nature of Silicon Dangling Bonds

Figure 4 shows two silicon surfaces imaged under the same conditions [14]. Both surfaces have a scattering of DBs. The left image is of a moderately n-type doped sample. It has been shown that DBs on such a surface are on average neutral. The DBs in that case are visible as white protrusions. The right hand image is of a relatively highly n-type doped sample. In that case each DB has a dark "halo" surrounding it. These DBs are negative. This results as the high concentration of electrons in the conduction band naturally "fall into" relatively low-lying DB surface state to make it fully, that is 2 electron, occupied. This localization of a negative charge at a DB causes destabilization of electron energy levels referred to as upward band bending. To a first approximation it is the inaccessibility of empty states for the STM tip to tunnel into that causes the highly local darkening of the STM image, that is, the halo. A much fuller description of the competing process involved in the imaging process have been described [15].

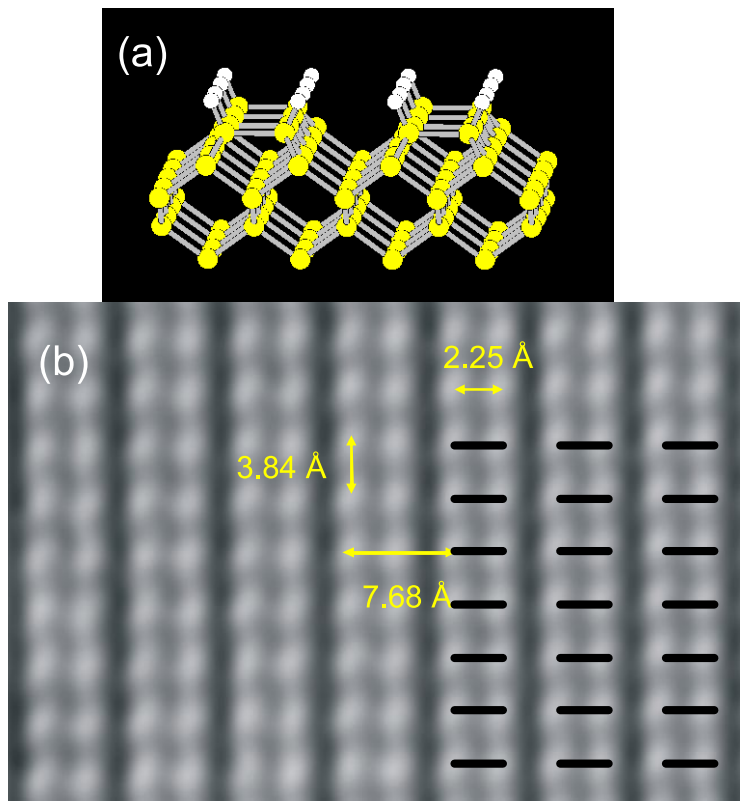

**Fig. 2.** (a) Model of a H-Si(100) surface with silicon atoms in yellow and hydrogen atoms in white. (b) Constant-current STM image of a H-Si(100) surface. Dimer rows are visible in the vertical direction and the atom separation along and across dimer rows are marked. Some dimer bonds are also marked by an overlaid grid of black bars.

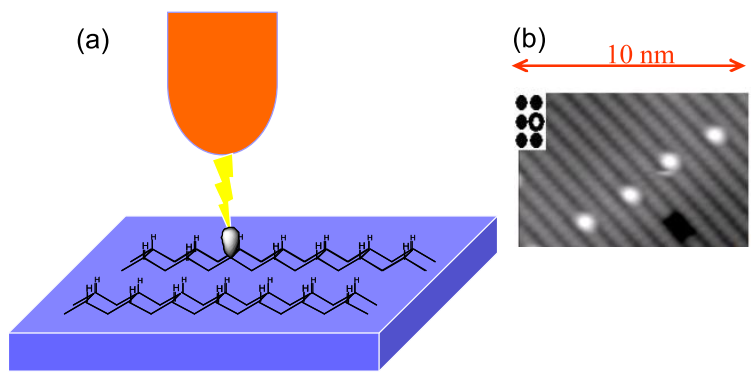

**Fig. 3.** Schematic of the fabrication of a dangling bond on the H-Si(100) surface by a scanned probe tip at a chosen location. (b) STM image of several DBs created in a line.

It is evident that the DB is effectively a dopant with a deep acceptor level. In accord with that character, a DB acts to compensate bulk n-type doping, causing the bands to shift up with respect to the Fermi level in the direction of a p-doped material. Most recently it has been shown that the neutral, single electron occupied DB can donate its charge to become positive thereby acting as a deep n-type dopant [16]. Summarizing, single electron occupation corresponds to neutral state. Two electron occupation corresponds to 1- charge. The absence of electrons in the DB leaves it in a 1+ charge state. The combination of dopant type, concentration, DB concentration on the surface, local electric field, and finally current directed through a DB, all contribute to determining its instantaneous charge state [15].

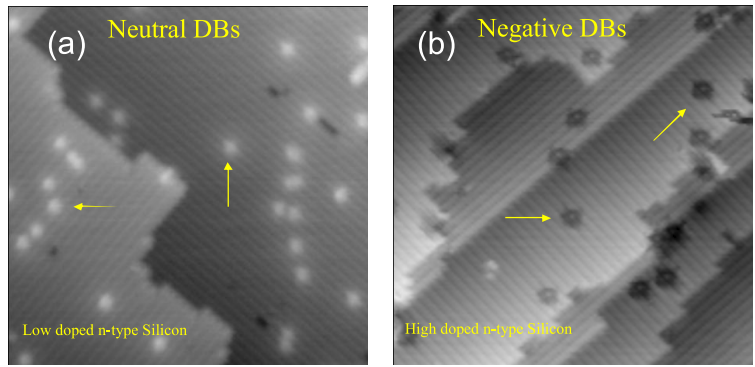

**Fig. 4.** Two silicon surfaces imaged under the same conditions. (a) A moderately n-type doped sample where DBs are on average neutral. (b) A highly doped sample where DBs are on average negatively charged.

A clean silicon surface, where every site has a dangling bond, is very reactive toward water, oxygen and unsaturated hydrocarbons like ethylene and benzene. While H atoms immediately react with a clean silicon surface, $H_2$ does not [17]. It is a remarkable fact that single DBs interact only weakly with most molecules, resulting in no attachment at room temperature. Most often, a second immediately adjacent DB is required in order for a molecule to become firmly bonded to the surface. Two DBs typically act together to form two strong bonds to an incoming molecule. This has the practical consequence that a protective layer can be formulated and applied to encapsulate and stabilizes DBs against environmental degradation.

A special class of molecules, typified by styrene, $C_8H_8$ attach to silicon via a self-directed, chain reaction growth mechanism [18]. As shown in Figure 5a, a terminal C reacts with a DB, thereby creating an unpaired electron at the adjacent C on the molecule. That species follows one of two paths. It either desorbs, or the radical C abstracts an H atom from an adjacent surface site to create a stably attached molecule, and a regenerated DB positioned one lattice step removed from the original DB position. The process repeats and repeats to create a multi-molecular line that gains a degree of order from the crystalline substrate.

Figure 5b shows an approximately 20 molecule long line of styrene grown in this way. The bright feature at the end of the line is a DB. It has been shown that under conditions where the terminal DB is negatively charged that charge acts to gate (Stark shift) the molecular energy levels causing conduction through the molecule where ordinarily it would not occur [19]. In other words the ensemble forms single-electron gated, one-molecule field effect transistor. The point of this discussion is to show there is precedent for microscopic observation of DBs at different charge-states, and to point out that a structure like the molecule transistor arrangement could be a useful detector of DB charge state [20].

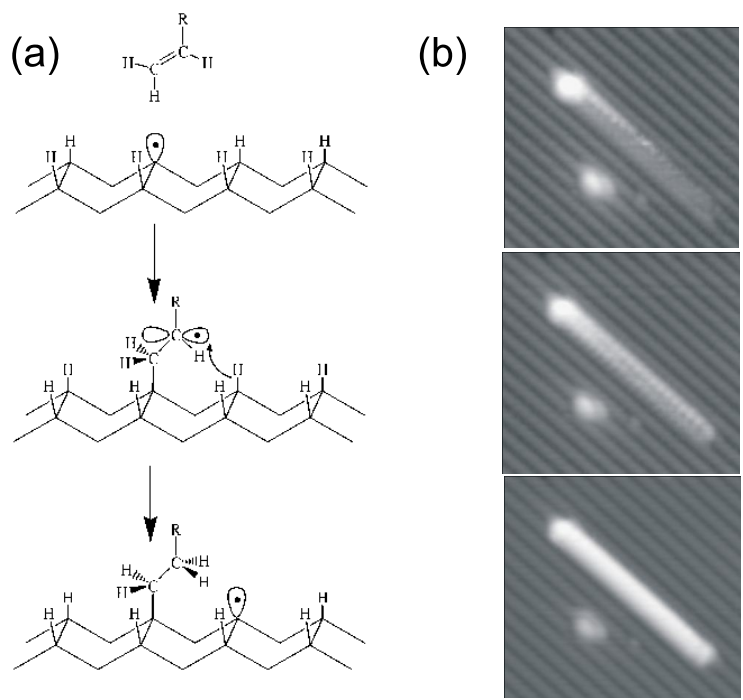

**Fig. 5.** (a) Various processes in a chain reaction resulting in the growth of lines a special class of molecules (e.g. styrene) on a H-Si(100) surface. R here denotes various possible radicals. (b) A line of about 20 styrene molecules grown in this fashion and imaged at different STM setpoints. The charge on the terminal DB at the top of the line depends on the STM setpoint and therefore causes (or not) a Stark shift in the molecular line.

## 4 Dangling bonds are Atomic Silicon Quantum Dots

The atoms in a silicon crystal enter into bonding and anti-bonding relationships with neighbouring and distant silicon atoms to form bands that span the crystal. In doing so the atoms give up their zero dimensional electronic character. Si atoms sharing in three ordinary Si-Si bonds and containing one dangling bond have a special mixed character. Like 4-coordinate silicon atoms, such atoms are very strongly bonded to the lattice and have an intimate role in the dispersive bands that delocalize electrons. At the same time, 3-coordinate atoms have one localized state, approximately of $sp^3$ character. This state is localized because it is in the middle of the band gap and mixes poorly with the valence and conduction band continua. The DB-containing atom is odder still in that the DB is partly directed toward the vacuum where it has a relatively limited spatial extent but is also partially contained within the silicon crystal where, because of dielectric immersion, is somewhat larger in spatial reach.

It was stated above that a DB state is like a deep dopant. Whereas a typical dopant has an ionization or affinity energy of several tens of meV, the DB has corresponding energies an order of magnitude larger. Consistent with that difference, the spatial extend of the DB state within the solid reaches several bond lengths, much less than the size of a common dopant atom [21].

The zero dimensional character of the DB, combined with the capacity to exhibit several (specifically 3) charge states leads us to think of the DB as a quantum dot. This may at first seem a bit odd as a quantum dot is often described as an artificial atom whereas we have a genuine atom, actually one part of an atom, forming our dot. But if a quantum dot is most fundamentally a vessel for containing and configuring electrons then, as subsequent examples will show, the ASiQD naturally and ably fits the definition, especially as the ease and precision of fabrication allows complex interactive ensembles of identical quantum dots to be made.

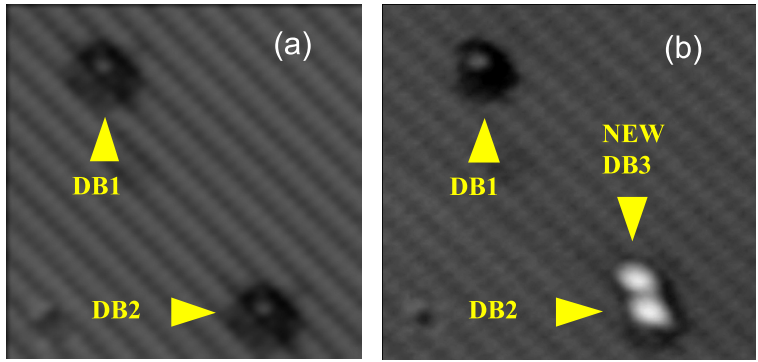

**Fig. 6.** (a) STM micrograph (10x10 nm, 2V, 0.2 nA) of a H-Si(100) surface with two DBs. The distinct dark halo indicates each DB is negatively charged. (b) An additional DB is created at a site near DB2 causing both the new DB3 and DB2 to appear very differently.

# 5 Fabricating and Controlling a Quantum Dot Cellular Automata Cell

Figure 6a shows two DBs. The distinct dark halo indicates the DBs are negatively charged. Figure 4b shows that when an additional DB is created by a tip directed H removal at a nearby site, both the new DB and the nearby pre-existing DB appear very differently, while the somewhat removed DB is unaltered. After extensive study it became clear that such a closely placed pair of DBs experiences a great Coulombic repulsive interaction, destabilizing the bound electrons and enabling one electron to leave the ensemble [14, 12]. The reduced net charge simultaneously stabilizes the remaining bound electron and creates an unoccupied energy level on one of the atoms. Because the barrier separating the DBs is low, of order several 100 meV, and is also very narrow, of order 2 nm, tunneling to the vacant state is very facile. Such a pair of DBs may be referred to as tunnel coupled. Our WKB and ab initio calculations agree that the tunneling rate for the 3.84 Angstrom separated DBs corresponds to an extremely short tunneling period of order 10 femtoseconds [22, 23]. Conventional relatively large and necessarily widely spaced dots would have a tunnel rate many orders of magnitude lower. Figure 7 shows the energy landscape schematically [14]. Each DB is represented by a potential well. The well is within the silicon bandgap. In figure 7a the separation between DBs is sufficiently large for the Coulombic interaction to be diminished by distance and by screening by conduction band electrons. In 7b the high energy repulsive relationship existing between two negatively charged DBs is represented. Figure 7b also shows the relaxed situation resulting after removal of one electron to the conduction band. In that final scenario one vacant electron state is shown. That state and the low and narrow barrier enables tunneling between the DBs.

The pairing result demonstrates a "self-biasing" effect. That is, by using fabrication geometry and repulsion to adjust electron filling, the need for capacitively coupled filling electrodes is removed [14]. Figure 8a shows several pairs of DBs of different separations and therefore different average net occupations. It can be readily seen that closer spaced DBs more fully reject one electron, leading to less local charge induced band bending and therefore to a lighter appearance in the STM image. The increasingly widely spaced pairs look increasingly dark as the net charge approaches 2 electrons. A statistical mechanical model of the paired DBs reproduces the effect as shown in Figure 8b. The graph stresses that occupation is a time averaged quantity and that pairs in the cross-over region will at any instant be either 1- or 2- charged [14].

Figure 9 shows a 4 dot ensemble or artificial molecule. The 4 dot cell was fabricated to result in an average net filling of 2 extra electrons. The graphs in 9b show the result of a statistical mechanical description of average occupation versus distance of separation in such a square cell at different temperatures [12].

One way to localize and thereby visualize the occupying electrons is to make an irregular shaped cell as is shown in Figure 10 [14]. Figure 10a shows three dots, two of which look darker indicating greater negative charge localization. Upon adding a fourth dot the previously darker sites become relatively light in appearance. This is due to the electrons attaining a lower energy configuration along a newly available longer diagonal. In a symmetric square or rectangular cell the freely tunneling electrons equally occupy the degenerate diagonal configurations. On the slow time scale of the STM measurement no instantaneous asymmetry can be seen.

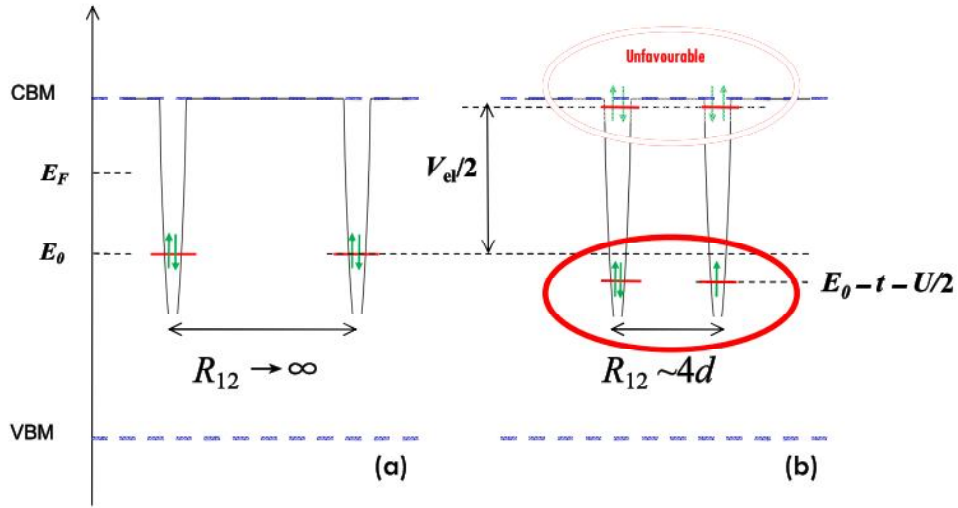

**Fig. 7.** The schematic energy landscape of a DB pair with each DB represented by a potential well with the ground state in the band gap. In (a), the separation between DBs is very large for the Coulombic interaction to be negligible and each DB is negatively changed (doubly occupied). In (b), DBs are much closer together (d is the dimer-dimer spacing) and a great Coulombic repulsion is associated with the doubly occupied configuration on both DBs. The diagram also shows the relaxed situation resulting after removal of one electron to the conduction band thus enabling tunneling of the remaining excess electron between the DBs.

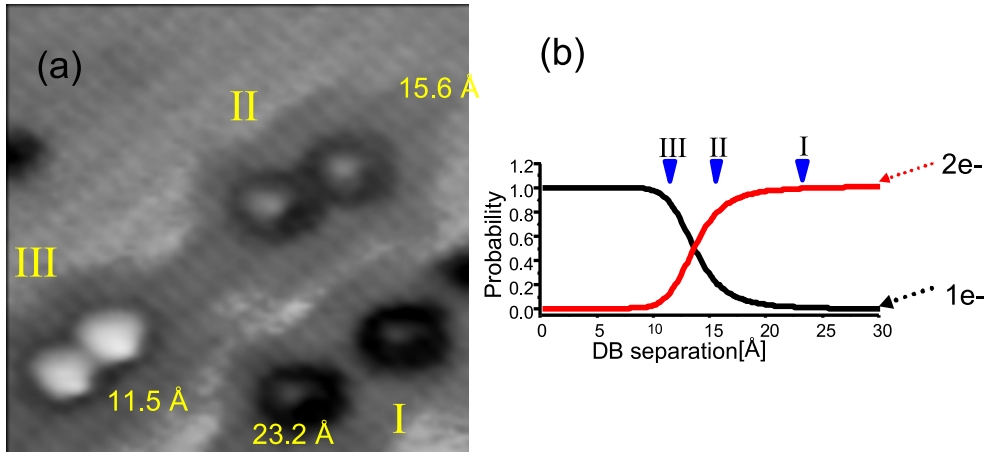

**Fig. 8.** (a) Several tunnel-coupled pairs of DBs fabricated at different separations (specified in each case) on the H-Si(100) surface. (b) Average occupation probability of a DB pair with 1 and 2 excess electrons as a function of DB separation. The three cases labeled in (a) are marked here with blue arrows.

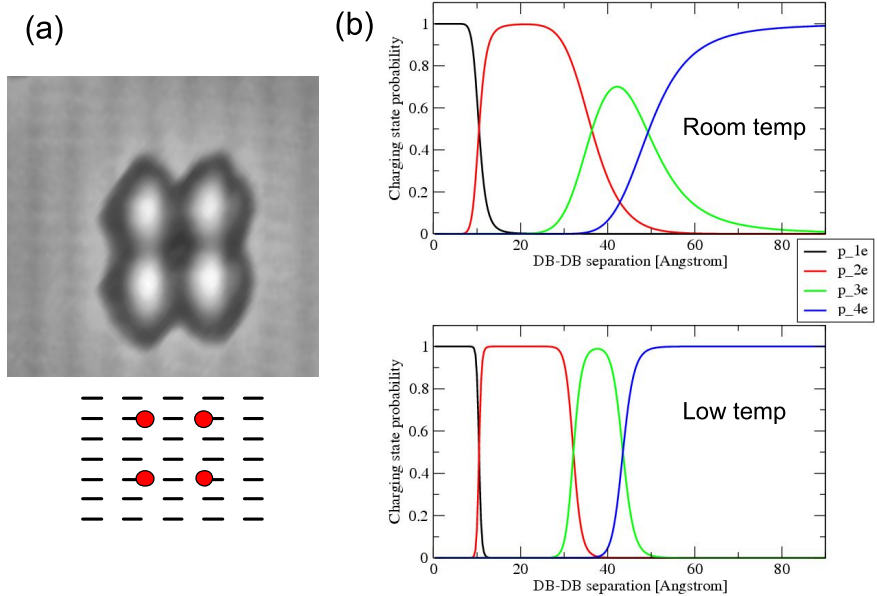

**Fig. 9.** (a) A fabricated ensemble (cell) of 4 tunnel-coupled DBs, or artificial molecule, calibrated to result in an average net filling of 2 extra electrons. A corresponding dimer lattice diagram is shown below. (b) The result of a statistical mechanical description of average occupation versus distance of separation in such a square cell at different temperatures (300 K top graph, 100 K bottom graph). The occupation probabilities with 1, 2, 3, 4 extra electrons are plotted for each case.

In order to embody the QCA architecture it must be possible to break that symmetry electrostatically and thereby to polarize electrons within a cell. This capacity is illustrated first by referral to a 2 dot cell. Figure 11 shows the sequential building of a 2 dot cell occupied by one extra electron and the polarization of that cell by one perturbing charge [14]. Figure 11a shows a small area, 3 nm across, of H-terminated silicon at room temperature. Figure 11b shows the creation of one ASiQD, while 11c shows the creation of a second ASiQD and the concomitant reduction in charge and darkness as seen by the STM. Upon charge removal, rapid tunnel exchange ensues. The coupled entity resulting may be described as an artificial homonuclear diatomic molecule. Like in an ordinary molecule, the Born-Oppenheimer approximation is valid. In other words, the electron resides so very briefly on one atom that nuclear relaxation does not have time to occur. On the electronic time scale, the nuclei are frozen. Finally in Figure 11d another charged DB is created. Using the knowledge displayed in Figure 6, the last DB is placed near enough to the molecule to affect it electrostatically, but not so close as to be tunnel coupled and a direct participant in the molecule. It is clear that the perturbation creates an analog of a heteronuclear diatomic molecule. The bond is polarized, it has ionic character. The perturbation controllably positions the electron. This is a key result. It shows that the type of coupling required between QCA cells, and between an electrode and a QCA cell is possible by this new approach. And all of this is possible at room temperature with an all silicon system.

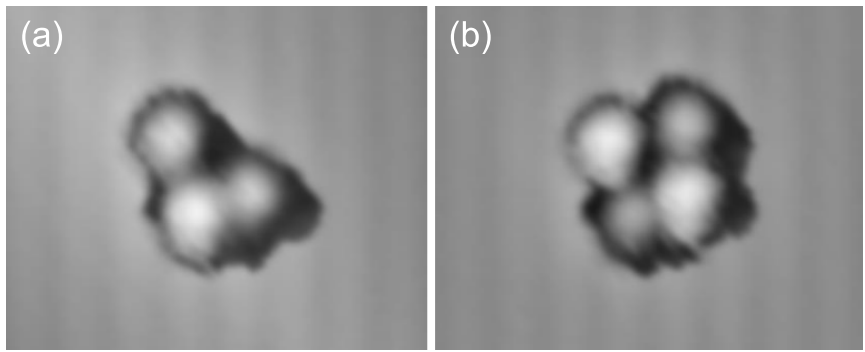

**Fig. 10.** (a) Three dots, two of which look darker indicating greater negative charge localization. (b) Upon adding a fourth dot in the upper right region the previously sites which previously looked dark become relatively light in appearance. This is due to the electrons attaining a lower energy configuration along the newly available longer diagonal.

Figure 12 shows the result of placing two perturbing electrons along one diagonal to place a 4 dot, 2 electron cell into one polarized binary state [14]. It is stressed again that this result was obtained at room temperature.

One could wonder about ways to increase the chemical stability of such complex QCA structures against environmental damage. As already discussed, for certain species such as styrene, there are known stable attachment mechanisms that require only a single DB. Such reactions can lead to molecular line growth on the surface. However the class of molecules that undergo such a process is rare. A passivating layer can be formed of molecules containing functional groups known not to react with a single DB. A yet broader class of molecules chemisorbs when two closely spaced DBs are provided. However, in order to cooperatively react with an incoming molecule, two DBs must be immediately adjacent, that is, co-located on the same underlying silicon dimer unit. Such DBs are separated by approximately 2.3 Angstroms. After reaction initiates at one of the two DBs, the second DB is just within reach by a C atom centered radical, allowing a second Si-C bond to form and resulting in a stable species and DB annihilation. When a second DB is not within reach, the single bonded species very quickly releases its grip because the interaction strength is only of order 0.1 eV. The DB in that circumstance is left unchanged. There is evidence of reactivity involving paired DBs separated by 3.84 Angstroms. The particular molecular functions able to form such a bond are readily avoided. In any case, none of the QCA patterns to be fabricated will include either of those reactive DB pairings. All of the DB cell structures will involve DBs spaced by approximately 8 Angstroms, or more.

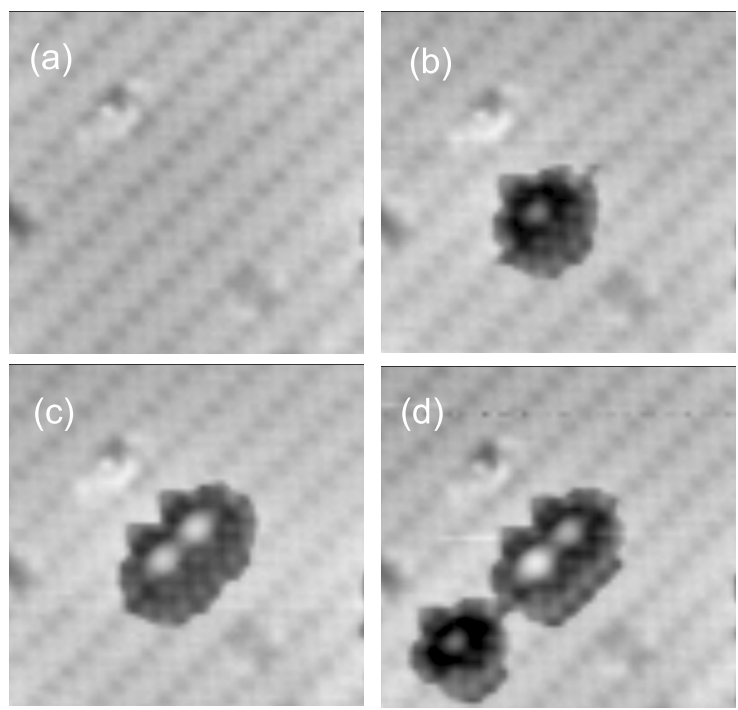

**Fig. 11.** The sequential building of a 2-dot cell occupied by one extra electron and the polarization of that cell by one perturbing charge. (a) A room temperature STM image approximately 3 nm across of H-terminated silicon surface. (b). The creation of one ASiQD. (c) The creation of a second ASiQD and the concomitant reduction in charge and darkness as seen by the STM. This tunnel coupled entity may be described as an artificial homonuclear diatomic molecule. (d) Another charged DB is created. This last dot is placed near enough to the molecule to affect it electrostatically, but not so close as to be tunnel coupled. It is seen that the perturbation creates an analog of a heteronuclear diatomic molecule as the bond is polarized.

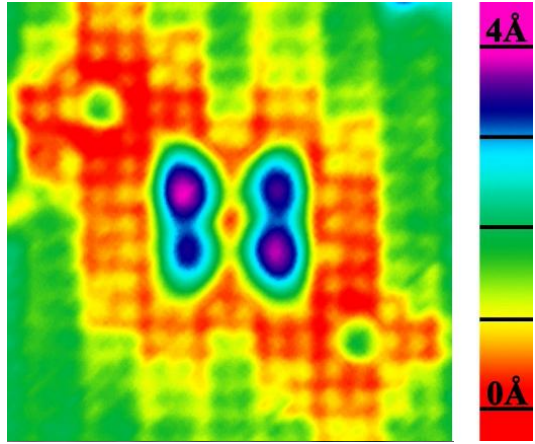

**Fig. 12.** Two perturbing electrons are positioned along one diagonal to place a 4-dot, 2-electron cell into one polarized binary state at room temperature.

## 6 The ASiQD as a New Element: Building Molecules

Our primary focus is on ultra low power, ultra fast field controlled computing as embodied by the QCA architecture or a variant thereof. We are well aware though of multiple other possibilities created by the ASiQDs. It is like having a new element to build with.

Figure 13 shows a collection of diverse assemblies or molecules [12]. The three-atom structures in (a) and (b) have slightly differently spacing. It is clear that different electronic structure results. The widely spaced grouping has distinct, less negative (brighter appearing) central atom while within the tightly spaced assembly all atoms appear the same. An in depth discussion is outside of our present scope. We will only note that the key difference between these structures is occupation. The wider spaced molecule tolerates a larger negative charge compared to the tightly spaced molecule. The latter has one electron. The widely spaced molecule has two electrons and those repel one another, leaving the central atom approximately neutral on average, and brighter in appearance. The complex arrangement labeled (d) uses both tunnel coupled atoms and perturbing atoms on the periphery to locally change the character of individual atoms and thereby the properties of whole ensemble.

Figure 13c shows a linear chain of ASiQDs. Such a structure will delocalize charge and allow biasing wires to be fabricated where needed. Such wires will bridge between relatively large lithographically created structures and the atom scale, allowing intimate input. It is a compelling feature of the ASiQD approach to circuit fabrication that high density ensembles enable passive components such as wires to be made while somewhat more widely spaced structures allow for the creation of the active elements in a circuit.

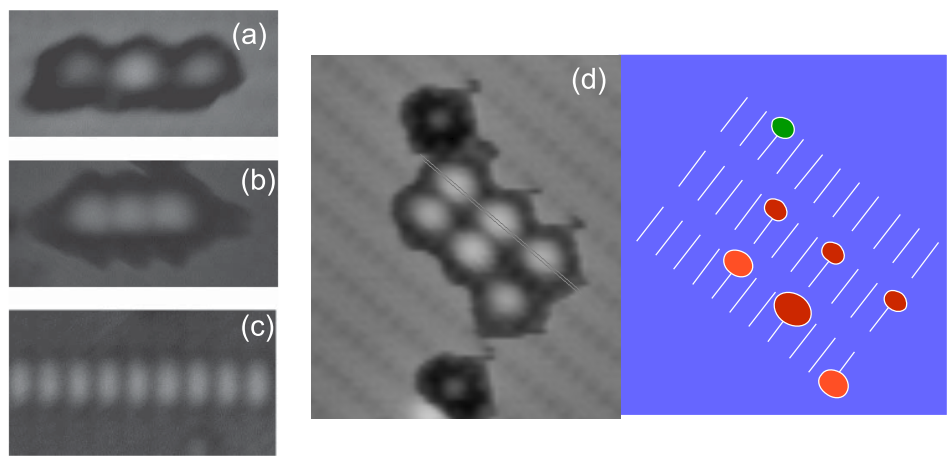

**Fig. 13.** A collection of diverse assemblies or molecules composed of atomic silicon quantum dots. The three-atom structures in (a) and (b) have slightly differently spacing. Different electronic structure results. The widely spaced grouping has a distinct, less negative (brighter appearing) central atom while within the tightly spaced assembly all atoms appear the same. The wider spaced molecule attains a larger negative charge compared to the tightly spaced molecule. The latter has one electron. The widely spaced molecule has two electrons and those repel one another, leaving the central atom approximately neutral on average, and brighter in appearance. (c) Is an image of a linear chain of ASiQDs. Such a structure will delocalize charge and allow biasing wires to be fabricated where needed. The complex arrangement labeled (d) uses both tunnel coupled atoms and perturbing atoms on the periphery to locally change the character of individual atoms and thereby the properties of whole ensemble. The perturbing atoms approximate the varied effects of different functional groups as used to tune organic molecular properties.

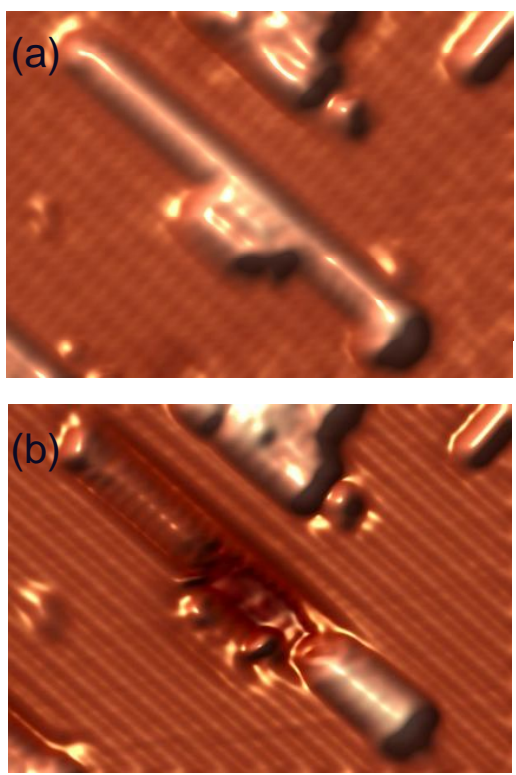

**Fig. 14.** STM images of single-triple $CF_3$-styrene / $OCH_3$-styrene heterostructure. (a) $V_s = 2V$ . (b) $V_s = -2V$. Single $OCH_3$-styrene and $CF_3$-styrene lines image above H-Si surface, while triple $CF_3$-styrene chains image below H-Si surface.

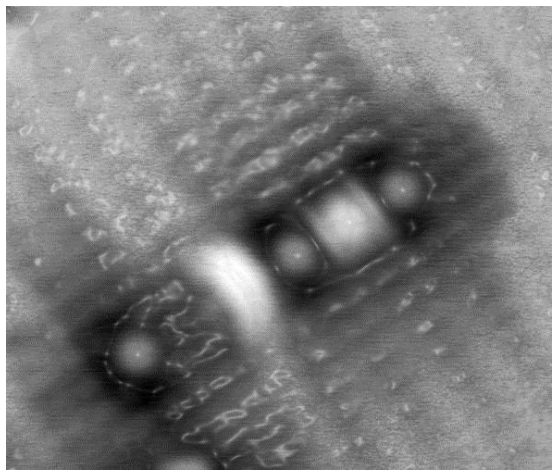

**Fig. 15.** A six atom molecule provides an example of the rich electronic structure within reach when molecules are designed and built in place where and as needed. The atoms are as close spaced as the lattice allows except between the 2nd and 3rd atoms counting from the left. Despite the uniformity of the constituents, the particular configuration results in a richly varied electronic property across the ensemble. Because the electronic structural maxima are not simply conformal to the positions of the constituent atoms it is not obvious without foreknowledge how many atoms are involved.

Molecules of designed optical properties can be made. This has been touched upon in a recent charge qubit characterization discussion [24]. More complex control is however possible. Not only the absorption energy, but the mode of adsorption can be pre-defined, whether electric dipolar, magnetic dipolar, or complex multipolar. The particular polarization dependence of absorption can be narrowly constrained too. Precise arrays of absorbers or emitters can be fabricated to achieve collective enhancements. Regular molecules and nano-clusters can also be used alone or in combination with ASiQD molecules to achieve designed function.

In addition to the perturbing effect of a nearby charge centre, attached molecules can be placed nearby to source a substantial and highly local multipolar field having the effect of making nearby ASiQDs more or less electron rich, rather like the approach of adding functional groups in chemistry. The pronounced effect of polarizing methoxy groups was experimentally and theoretically described in [25, 26] and is shown in Figure 14.

Finally we show as Figure 15 an example of the kind of rich electronic structure within reach. In this case 6 ASiQDs were prepared in a line. The atoms are as close spaced as the lattice allows except between the 2nd and 3rd atoms counting from the left. Because the electronic structural maxima are not simply conformal to the positions of the constituent atoms it is not obvious without foreknowledge how many atoms are involved. A catalogue of such molecules is being prepared in concert with modeling to extend known structure-property relationships.

For over two decades molecular electronics research has sought to transfer molecules from a bottle to a surface to achieve desirable circuitry. In practice, studied configurations between molecules and electrodes are uncertain and variable, as are the properties of the ensemble. While progress toward self-assembling molecular structures has been made, for the most part the linking forces directing assembly of molecules have been focused on geometric arrangements without consideration or control of electrical connectivity. In some instances when single molecules have been rendered to conduct electrical current, molecular decomposition has promptly occurred. We are considering initiating what could be considered as "phase 2" of molecular electronics. By building artificial molecules where and as needed, and by precisely building contacts also, it is possible to ensure interactions and properties are exactly as designed. Working only with the limited palette described here, it is evident that artificial molecules of diverse properties can be designed and made in place. Networks of high complexity and function can be formed. At the cost of greater fabrication complexity, for example through the addition of atoms other than H and Si, a far broader range of properties will be available. It is anticipated that efficient substrate coupling will diminish decomposition pathways and that artificial molecules as envisioned here will not fail as organics will as a result of undesired excitation and heating.

# 7 Instrumentation and Custom Lithography to Make Prototype QCA Circuitry

State of the art instrumentation is required to make advances in this area. Years of ordinary STM investigations were hampered by at least two problems. One is non-ideal scanning and fabrication control  something we will refer again to in the next section. The other aspect is a lack of a bridge between the atom sale and the macro scale.

Our first approach to making sufficiently fine lithographic features to controllably interact with atom scale structures began nearly 20 years ago. Titanium silicide contacts were prepared using a normal optical lithography and lift-off approach [27, 28]. When examined at the atomic scale, lithographic features prepared in this way are unacceptably rough and crudely defined for our purposes. The transition from pure silicide to pure silicon is not abrupt, but spans 10s of nm, and is of unknown composition and of uncontrolled electronic character. However, it is possible to grow crystalline Titanium Silicide features on a silicon substrate with atomically-precise boundaries by simply evaporating titanium and annealing in ultra high vacuum. Figure 16 shows a highly crystalline Titanium Silicide island on a silicon substrate. The boundary is atomically abrupt. The silicon nearby is well ordered as required. No spatial patterning is imposed. Refinements of this technique will enable the type and quality of contacts re-quired. Ultra fine extensions from such contacts with linear wire structures composed of closely spaced ASiQDs will enable precise electrostatic addressing of atomic structures.

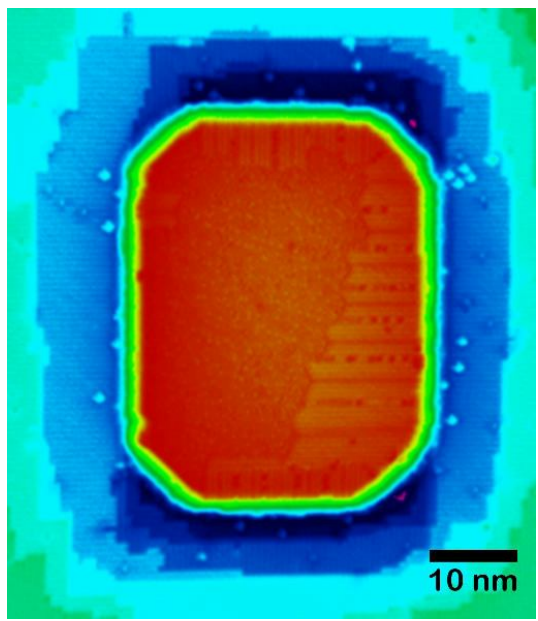

**Fig. 16.** A crystalline Titanium Silicide island on a silicon substrate. Crucially, the silicide-silicon boundary is atomically abrupt. The silicon nearby is well ordered as required. Ultra fine extensions from such contacts with linear wire structures composed of closely spaced ASiQDs will enable precise electrostatic addressing of atomic structures.

The potential to controllably alter the charge state of individual DBs has been demonstrated [29]. Figure 17 shows a schematic diagram of a silicide contact and nearby DBs. Because of the spatially varying potential built-in at the silicide-silicon interface (and because the H-terminated surface is not pinned) the DBs nearby take position dependent charge states. Figure 17b shows an experimental verification of the scheme. Going forward, combined lithographic approaches will allow for suitably small and high quality lithographic features. Multiple contacts will be connected and active while a device is in the STM fabrication and inspection tool allowing prototyping methods and device testing to advance substantially over what has been available to date.

In parallel with the development of nano-lithographic methods, a multi-probe STM has been developed to allow nano-scale electrical characterization that has until now been out of reach. The instrument shown in Figure 18 has three independently scannable tips, watched over by a scanning electron microscope. Each tip can be quickly redeployed as a scanned probe for imaging or touched down as a current source or as a voltage probe. Initial applications have enabled the first detection of a potential step on a crystalline silicon surface [30], the absolute measurement of the Si(111)-7x7 surface conductivity [31], the absolute value of the Si(111)-7x7 step resistivity [31] and the conductivity of a DB wire on the H-Si(100) surface [32]. Various measurements related to and enabling of atomic silicon surface electronic circuitry will follow.

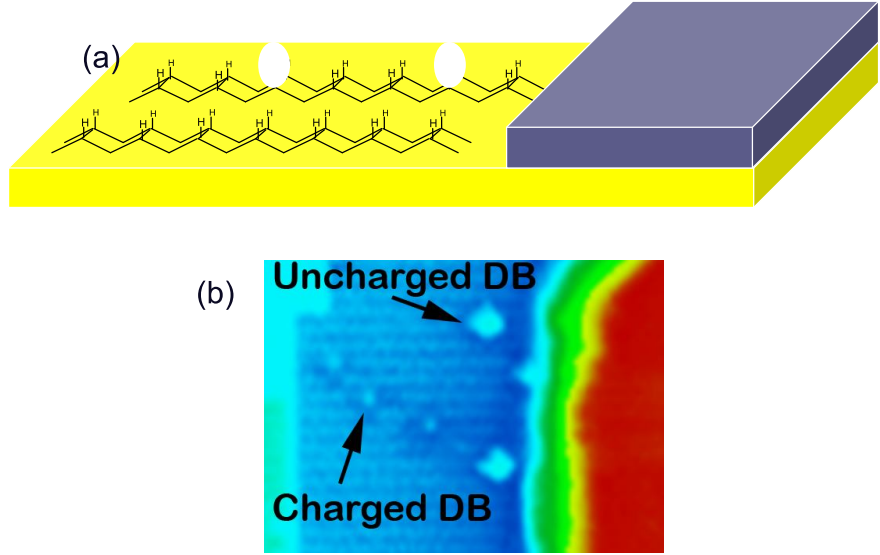

**Fig. 17.** A schematic diagram of a silicide contact and nearby DBs. The spatially varying potential built-in at the silicide-silicon interface causes DBs nearby to take position dependent charge states. (b) An experimental verification of the scheme.

## 8 Quantum Silicon Incorporated and First Proto-Circuits

A spin-off company has been created to exploit the recently discovered DB/ASiQD properties reviewed here. The goal is to undertake focused development of all the many components needed to demonstrate working atom scale field controlled computing elements. In so doing, fundamental components and methods will be developed that also enable analog and quantum circuitry.

It has been 2 decades since Lyding, Tucker and associates first made DB wires [33]. Though many subsequent studies explored other aspects of that system limited progress has been made in fabricating atomically precise patterns. Making precise DB patterns is difficult with standard equipment. The first ASiQD paper several years ago made the key step forward [14]. Despite years of work by many researchers the evidence of coupling among DBs and their quantum dot character was a surprising and disruptive event. It changed entirely what was known to be possible with DBs.

There were numerous other difficulties slowing the recognition of DBs as circuit building blocks. Though dangling bond type defects were identified by Bardeen in the 1940s only in the last several years has a robust theory describing the properties of the DBs and importantly how those are manifest in STM images been completed [12, 15, 34–36]. While clear indications that the DB can exhibit a positive charged state were recorded by us one decade ago those results went unreported for want of fuller theoretical or experimental verification. Additional experimental proof has appeared just in the past year and been reported together with the original observations [16].

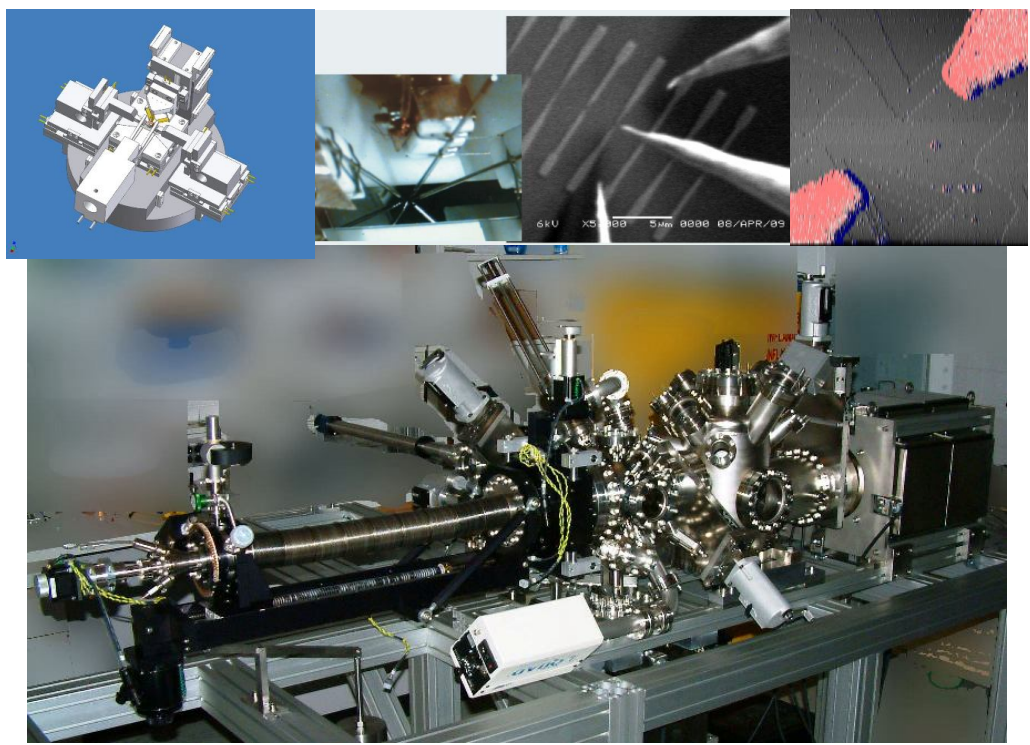

**Fig. 18.** A multi-probe STM has been developed to allow nano-scale electrical characterization that had previously been beyond reach. The instrument has three independently scannable tips, watched over by a scanning electron microscope. Each tip can be deployed as a scanned probe for imaging or touched down as a current source or as a voltage probe. Nanoscale potentiometry and transport measurements have been made with the instrument.

While the above advances have established the value of DBs as circuit elements it has remained clear that wished-for ensembles of DBs would never be studied or deployed until advances in fabrication were made. The 2006 ASiQD paper reported on rare successful fabrication efforts. It was then impractical to make even a 2 cell QCA structure [13].

Efforts to better understand and control STM scanning have led to some improvements [37, 38] but fell short of control necessary to fabricate atom scale QCA circuitry. Recently, our reevaluation of non-idealities inherent to the scanned probe fabrication process and in the character of the scanned probe tip itself have led to a large improvement. Yields of a tiny fraction of 1% have jumped to 80%. The various refinements will not be discussed here but the results can be seen in Figure 19. A 16 atom, 4 QCA cell is shown with one atom out of place. Further study has uncovered the dominant reasons for this remnant fabrication error and it is anticipated that as new fabrication tools come on line in the year ahead that improved yields will result. As a test of larger pattern fabrication capacity, and despite the errors still produced, two well studied QCA circuits from the literature [39] were made. We see good overall pattern fidelity and a clear demonstration that we have broken free of the 4 atom limit of a few years ago and may soon be able to make 100 atom structures with excellent fidelity. The circuits in Figure 19 required about 1 minute to fabricate and were automatically made by computer upon input of pattern required.

Substrate defect elimination, automatic fabrication error correction, as-needed on the fly rerouting and circuit redundancy strategies are all parts of the plan to achieve functional devices within the next several years. Numerous contact, input/output, encapsulation and packaging issues among others which we have charted courses for will not be discussed in this document. Likewise approaches to eventual very rapid parallel fabrication processes are being developed but are beyond the present scope of discussion.

## 9  ASiQDs for Quantum Computing

In the QCA mode of operation and in analog electronic strategies the resulting device is not quantum in that it does not depend upon quantum coherence or superposition. However, it is anticipated that ASiQD-based qubits for eventual quantum computing applications can be made - both charge-based and electron spin-based qubits are possible.

The key attribute of the ASiQD-based charge qubit is that the rate of tunneling is very large [23, 24]. It is estimated that the tunnel rate can be as much as 106 times larger than the rate of decohering events. Practically this means that there could be sufficient time to undertake a coherent operation before a disruption occurs. Though this is an attractive situation, it is also challenging as it means that precisely phase-controlled operations on the qubit must be done with fraction tunnel period time control which is not possible with any conventional electronics. Some form of optical control is suggested therefore. A new strategy for characterization of the charge qubit has been presented [24].

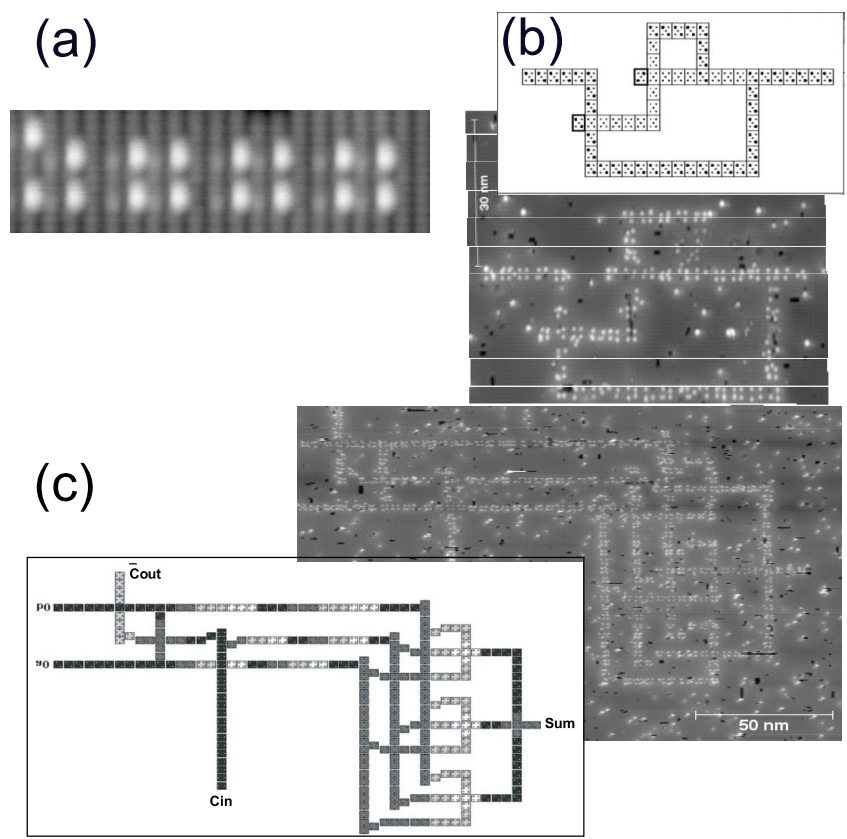

**Fig. 19.** (a) A 16 atom, 4 QCA cell is shown with one atom out of place, demonstrating the near perfection, but occasional errors resulting at this time. (b), (c) Larger pattern fabrication capacity demonstration. The circuits required about 1 minute to fabricate and were automatically made by computer upon input of pattern required.

The spin-type qubit would minimally use one ASiQD per qubit. Unlike in all of the above discussion where it was presumed the isolated DB would have a 1-charge, the DB would be prepared in a neutral one-electron state. Application of a magnetic field differentiates the up and down spin states to create a suitable Zeeman-split two-level system. This is similar to the P atom dopant approach. There, the P atom is studied at very low temperature where it does not ionize and so is not in fact a dopant. It retains its one extra electron. The resulting paramagnetic centre has desirable electron spin properties as proven by ESR measurements of very large numbers of P atoms.

An architecture for spin-based quantum computing using P dopants was first put forward by Kane [40]. In that scheme, the nuclear spins of P atoms and the spins of bound electrons serve as qubits, which interact via hyperfine- and exchange-interactions. In pursuit of that goal, great effort has been devoted over the last decade to place single P dopant atoms into a silicon lattice either by a chemical process that achieves nm position control but not perfect atom scale control, or by ion implantation which is highly uncontrolled and not scalable but which has led to the most impressive results so far [41–43]. The fabrication challenges facing P dopant approaches to quantum computation are significant, and it is known that quantum computation cannot work even in principle until these are overcome [44] A later proposal by Loss and DiVincenzo [45] makes possible an all-electron-spin approach to quantum computation, which could do away with the need for the nuclear spins of P donors. Paramagnetic ASiQDs are a suitable platform for this architecture or some variant of it, and immediately offer the advantage of atomically-precise fabrication, in addition to the important features of electronic control and silicon compatibility.

Presented in this way, the DB appears to be a far more attractive electron-spin qubit than the implanted P atom in silicon. That is our belief. But this writing is the first to our knowledge to point out the many advantages. The main disadvantage of the DB route to spin-based quantum computing is that passivation or encapsulation is required, but that seems a surmountable problem. The advantages to using DBs are many. Unlike P atom insertion through a multi-step process, DBs can be made instantly. While P atoms cannot be placed exactly and reproducibly with respect to other P atoms, any number of ASiQDs can be perfectly juxtaposed, just as designed. Some of the latest strategies for achieving robust qubits by combining multiple physical qubits into one logical qubit [46] are greatly aided by this precise multi qubit fabrication facility.

The DB has another advantage related to its large ionization energies. First, the paramagnet species easily exists at room temperature. While low temperature will still be required to avoid lattice-induced decoherence effects, the deeply held electron in the DB will be advantageous over the relatively weakly held P electron. An important advantage emerges trivially because of the relative electronic sizes of the DB and P atoms. Being small and well confined, the DB interacts with fewer surrounding nuclei than the P atom which encompasses an order of magnitude larger volume. The DB as a result will experience fewer nuclear spin-electron spin decoherence effects.

## 10 Conclusions

In this paper, we outlined our recent and current efforts in building atom-scale quantum-dot cellular automata circuits on a silicon surface. As a building block we use the silicon dangling bond on a H-Si(001) surface, shown to act as a quantum dot. The fabrication, experimental STM imaging, and charging characteristics of the dangling bond and their assemblies are discussed. We then show how precise assemblies of such dots can be created to form artificial molecules. Such complex structures can be used as systems with custom optical properties, circuit elements for quantum-dot cellular automata, and quantum computing. Considerations on macro-to-atom connections are discussed.